# Neighborhood-Based Label Propagation in Large Protein Graphs

Sabeur Aridhi [1*], Seyed Ziaeddin Alborzi [2], Malika Smaïl-Tabbone [1], Marie-Dominique Devignes [3], David W Ritchie [2]

[1] University of Lorraine, LORIA, BP 239, 54506 Vandoeuvre-lès-Nancy, France
[2] INRIA Nancy Grand Est, 54600 Villers-lès-Nancy, France
[3] CNRS, LORIA, BP 239, 54506 Vandoeuvre-lès-Nancy, France
*To whom correspondence should be addressed: sabeur.aridhi@loria.fr

1. INTRODUCTION

Understanding protein function is one of the keys to understanding life at the molecular level. It is also important in several scenarios including human disease and drug discovery (1). In this age of rapid and affordable biological sequencing, the number of sequences accumulating in databases is rising with an increasing rate (2). This presents many challenges for biologists and computer scientists alike. In order to make sense of this huge quantity of data, these sequences should be annotated with functional properties. UniProtKB consists of two components: i) the UniProtKB/Swiss-Prot database containing protein sequences with reliable information manually reviewed by expert bio-curators (3,4) and ii) the UniProtKB/TrEMBL database that is used for storing and processing the unknown sequences (4). Hence, for all proteins we have available the sequence along with few more information such as the taxon and some structural domains (identified on the protein 3D structure or reliably predicted from the primary sequence). Pairwise similarity can be defined and computed on proteins based on such attributes. Other important attributes, while present for proteins in Swiss-Prot, are often missing for proteins in TrEMBL, such as their function and cellular localization. The enormous number of protein sequences now in TrEMBL calls for rapid procedures to annotate them automatically.

2. LABEL PROPAGATION IN LARGE PROTEIN GRAPHS

Here, we present DistNBLP, a novel **Dist**ributed **N**eighborhood-**B**ased **L**abel **P**ropagation approach for large-scale annotation of proteins. To do this, the functional annotations of reviewed proteins are used to predict those of non-reviewed proteins using label propagation on a graph representation of the protein database. DistNBLP is built on top of the "akka" toolkit for building resilient distributed message-driven applications using the nodes of a physically independent network of machines (5). DistNBLP takes as input a graph representation of the protein data. Each node of the graph represents a protein while an edge between two nodes means that the linked proteins exhibit a minimum similarity (of any kind). Each node $i$ is identified by a set of labels $L(i)$ (one or more annotations to propagate), has a set of neighbors $N(i)$ and for every neighbor $j$ an associated weight $W_{ij}$. DistNBLP works in two main steps. The first step consists of partitioning the input graph into multiple connected subgraphs by a partitioner, each subgraph being assigned to a different worker. Each worker is aware of all the neighbors of its subgraph nodes in distant workers. The partitioner supports several types of predefined partitioning techniques. The overall algorithm is coordinated by a master which orchestrates communication between the workers. The second step consists of distributed propagation of protein labels. Each worker loads its subgraph and performs local computations (involving communication with distant workers), after which the status of the nodes in every subgraph is updated according to Algorithm 1.

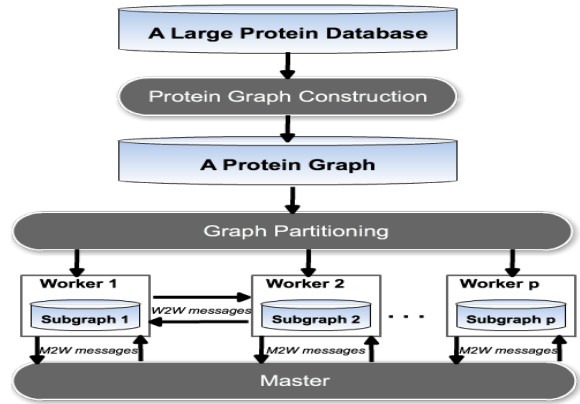

Figure 1: System overview of DistNBLP

| Algorithm 1: DistNBLP |
|---|
| **Repeat** |
|     **Foreach** node *i* (vertex of the graph) **do** |
|         Broadcast the *set of labels* of *i* (*L(i)*) along with the weight $W_{ij}$ (similarity values) to each neighbor *j* (including neighbors present in other workers) |
|         **If** *L(i)* is empty **then** |
|             Sum the weights (similarity values) for each collected label |
|             Update *L(i)* with the majority label |
|         **Else** Discard the received labels |
|     **End Loop** |
| **Until** all the nodes are labelled |

At each iteration, multiple workers execute in parallel. Within each worker operation, the states of its nodes are updated according to Algorithm 1. DistNBLP uses two messaging modes: i) MasterToWorker (M2W)/WorkerToMaster (W2M) that allows message exchanges between the master and the workers, and ii) WorkerToWorker (W2W) that allows message exchanges only between workers. The W2W mode is used to broadcast the label of a node *i* to another node *j* when *i* and *j* are not processed by the same worker. The M2W/W2M mode is used to check at each iteration if the workers have finished. The master stops the algorithm when the stopping criteria is verified (all of the nodes of all the workers have been labelled). We used DistNBLP in the CAFA3 (Critical Assessment of Functional Annotation) experiment in which the aim was to annotate around 130,787 protein sequences with GO molecular function (MF) terms. Using the number of shared Pfam domains as a measure of pair-wise protein similarity in the constructed protein similarity graph, DistNBLP predicted a total of 345,271 GO MF annotations for 99,625 protein sequences. The results are under evaluation.